\documentstyle[preprint,aps]{revtex}
\tightenlines 

\begin{document} 

\title{Teleportation Criteria: Form and Significance}
\author{T.C.Ralph}
\address{Department of Physics, Centre for Lasers, \\
University of Queensland, QLD 4072, Australia\\
E-mail: ralph@physics.uq.edu.au}

\maketitle

\begin{abstract}
Our criteria for continuous variable quantum teleportation [T.C.Ralph 
and P.K.Lam, Phys.Rev.Lett. {\bf 81}, 5668 (1998)] take the form of sums, rather 
than products, of conjugate quadrature measurements of the signal 
transfer coefficients and the covariances between the input and output 
states.  We discuss why they have this form. We also discuss the 
physical significance of the covariance inequality.
\end{abstract}

\vspace{8mm}

Recently we have proposed criteria for the characterization of 
continuous variable quantum teleportation \cite{ral98}.  It was shown 
that for any classical teleportation scheme (i.e. where no 
entanglement is shared) the following inequalities cannot be violated 
for minimum uncertainty Gaussian input states:
\begin{eqnarray}
T_{t} & = & T_{s}^{+}+T_{s}^{-} \le 1
\label{crit1} \\
V_{t} & = & {{1}\over{2}}(V_{cv}^{+}+V_{cv}^{-}) \ge 1
\label{crit2}
\end{eqnarray}
Here $T_{s}^{+}$ ($T_{s}^{-}$) are the amplitude (phase) quadrature signal
transfer functions from the input to output fields of the teleporter
defined by
\begin{equation}
T_{s}^{\pm}={{{\rm SNR}_{\rm out}^{\pm}}\over{{\rm SNR}_{\rm in}^{\pm}}}
\end{equation}
where the ${\rm SNR}$ are the signal to noise ratios of small classical
test signals. The RF frequency test signals are placed
on each quadrature of the input
beam. The conditional variances between the input and output for each
of the quadrature amplitudes are defined by
\begin{equation}
V_{cv}^{\pm}(\omega)=V^{\pm}_{\rm out}(\omega)(1-C^{\pm}(\omega))
\end{equation}
where $V_{\rm out}^{\pm}(\omega)= \langle |\delta X_{\rm 
out}^{\pm}(\omega)|^{2}\rangle $ are the standard spectral variances for the 
amplitude ($\delta X^{+}$) and phase ($\delta X^{-}$) quadrature 
fluctuations of the output state assessed at some RF frequency 
$\omega$.  The correlation function, $C$, is defined by
\begin{equation}
C^{\pm}={{|\langle \delta X_{\rm in}^{\pm}
\delta X_{\rm out}^{\pm}\rangle|^{2}}\over{V_{\rm in}^{\pm}V_{\rm out}^{\pm}}}
\end{equation}
and is directly related to the ${\rm SNR}$'s via $C^{\pm}=T_{s}^{\pm}$ 
\cite{poi94} when cross-coupling between the quadratures can be 
ignored \cite{note1}.  The criteria in Eqs.~(\ref{crit1}) and 
(\ref{crit2}) are semi-independent.  Violation of either inequality 
indicates entanglement is present.  A strong test of teleportation 
would require both inequalities be violated simultaneously.

Although there is no question of the validity of the inequalities in 
Eqs.~(\ref{crit1}) and (\ref{crit2}) there has been some discussion 
about their significance \cite{loo99}. 
In particular, it has been queried as to 
why sums are formed of the conjugate quadrature measurements instead 
of the ``usual'' procedure of forming products (cf: Heisenberg 
uncertainty principle).  In the following we justify the form of 
$T_{t}$ and $V_{t}$ by deriving them from more fundamental 
expressions.

Suppose the quantum fluctuations of our input field are described in 
the usual way by the zero-mean
annihilation operator $\delta a_{\rm in}$ whilst our output field is 
described by $\delta a_{\rm out}$.  The amplitude and phase quadrature 
fluctuations of the input field are defined respectively by $\delta 
X_{\rm in}^{+}=\delta a_{in}+\delta a_{in}^{\dagger}$ and $\delta 
X_{\rm in}^{-}=i(\delta a_{in}^{\dagger}-\delta a_{in})$.  The quadrature 
fluctuations of the output field can be described by
\begin{equation}
\delta X_{\rm out}^{\pm}=Y^{\pm}\delta X_{\rm in}^{\pm}+Z^{\pm} \delta X_{N}^{\pm}
\label{x}
\end{equation}
where $Y^{\pm}$ and $Z^{\pm}$ are c-numbers and $\delta X_{N}^{\pm}$ includes
all added noise sources. Here again we will assume there is no cross
coupling between the amplitude and phase quadratures of the input and
output \cite{poi94},\cite{note1}. Then
\begin{equation}
V_{\rm out}^{\pm}=|Y^{\pm}|^{2} V_{\rm in}^{\pm}+|Z^{\pm}|^{2} V_{N}^{\pm}
\label{v}
\end{equation}
and
\begin{equation}
T_{s}^{\pm}={{V_{\rm in}^{\pm}|Y^{\pm}|^{2}}\over{|Y^{\pm}|^{2} V_{\rm in}^{\pm}+
|Z^{\pm}|^{2} V_{N}^{\pm}}}
\label{ts}
\end{equation}
Because the information used to produce the output field traveled 
through a classical channel the added noise 
terms must be sufficient to ensure the generalized uncertainty 
principle is maintained for the photo-currents in the classical 
channels \cite{poi94},\cite{art88}, this implies
\begin{equation}
{{|Z^{+}|^{2}}\over{|Y^{+}|^{2}}} V_{N}^{+} {{|Z^{-}|^{2}}\over{|Y^{-}|^{2}}}
V_{N}^{-} \ge 1
\label{noise}
\end{equation}
For a minimum uncertainty input state ($V_{\rm in}^{+}V_{\rm in}^{-}=1$)
and using Eq.~(\ref{ts}) the inequality in Eq.~(\ref{noise}) reduces to that
of Eq.~(\ref{crit1}). Thus we see that the origin of the signal transfer
inequality is, in fact, the standard product uncertainty. In quantum
teleportation the shared entanglement acts as a ``quantum key''
that enables the inaccessible (because of the added
noise) quantum information on the classical channel to be retrieved
\cite{ral98,deu99}.
The noise on the output field can then be less than the minimum
required for the classical channel thus allowing the inequality to be
violated.

We now consider how one might generalize the conditional variance used 
in QND \cite{poi94,hol90} to teleportation.  
In QND one is only interested in how well the properties of one 
quadrature are preserved.  In teleportation we wish to quantify how 
well the entire state is preserved.  As we are working in the 
Heisenberg picture, in which it is the operators which evolve in time, 
this is equivalent to quantifying how well the field operator, $a$, is 
preserved.  Caves \cite{cav82} defined the variance of the field 
operator as
\begin{equation}
V_{f}={{1}\over{2}} \langle \delta a^{\dagger}\delta a+\delta a \delta 
a^{\dagger} \rangle
\end{equation}
Following the standard procedure for constructing a conditional
variance we are thus motivated to consider the field correlation
\begin{equation}
C_{f}={{|{{1}\over{2}} \langle \delta a_{\rm out}^{\dagger}\delta 
a_{\rm in}+\delta a_{\rm in} \delta a_{\rm out}^{\dagger}+\delta a_{\rm out}\delta 
a_{\rm in}^{\dagger}+ \delta a_{\rm in}^{\dagger} \delta a_{\rm out} \rangle 
|^{2}} \over{\langle \delta a_{\rm out}^{\dagger}\delta a_{\rm out}+ \delta 
a_{\rm out} \delta a_{\rm out}^{\dagger}\rangle \langle \delta 
a_{\rm in}^{\dagger}\delta a_{\rm in}+\delta a_{\rm in} \delta a_{\rm in}^{\dagger} 
\rangle}}
\end{equation}
If the fields are identical, i.e. $\delta a_{\rm in}=\delta a_{\rm out}$,
then $C_{f}=1$. If the fields are completely independent, i.e.
$[a_{\rm out},a_{\rm in}^{\dagger}]=[a_{\rm in},a_{\rm out}^{\dagger}]=0$,
then $C_{f}=0$. We then construct the
field conditional variance as
\begin{equation}
V_{cv f}= \langle\delta a_{\rm out}^{\dagger}\delta a_{\rm out}+
\delta a_{\rm out} \delta  a_{\rm out}^{\dagger}\rangle(1-C_{f})
\label{cf}
\end{equation}
Using $\delta a={{1}\over{2}}(\delta X^{+}+i \delta X^{-})$, we can
rewrite Eq.~(\ref{cf}) as
\begin{eqnarray}
V_{cv f} & = & {{1}\over{2}}(V_{\rm out}^{+}+V_{\rm out}^{-}-{{|\langle\delta
X_{\rm out}^{+} \delta X_{\rm in}^{+}+\delta
X_{\rm out}^{-} \delta X_{\rm in}^{-}\rangle |^{2}}\over{V_{\rm 
in}^{+}+V_{\rm in}^{-}}}) \nonumber\\
 & = & {{1}\over{2}}(V_{\rm out}^{+}+V_{\rm out}^{-}-
 {{(\sqrt{T_{s}^{+}V_{\rm out}^{+}V_{\rm in}^{+}}+
 \sqrt{T_{s}^{-}V_{\rm out}^{-}V_{\rm in}^{-}} )^{2}}\over{V_{\rm 
in}^{+}+V_{\rm in}^{-}}})
\label{cf2}
\end{eqnarray}
For independent fields $V_{cv f}\ge 1$. If we assume that 
$Y^{+}=Y^{-}$, i.e. the teleporter acts symmetrically on the two 
quadratures of the input field, then it is straightforward to show 
that in fact
\begin{equation}
V_{cv f}=V_{t}
\end{equation}
Thus the inability of classical teleportation schemes to violate the
inequality of Eq.~(\ref{crit2}) can be seen as showing that
in a certain sense the input and output remain independent fields.
The sum form of the 
inequality is seen to have its origin in the structure of the field 
operator as a sum of the two quadrature components. For asymmetric 
manipulations of the input quadratures ($Y^{+}\ne Y^{-}$) we find 
$V_{cv f}\ne V_{t}$. Our analysis suggests that $V_{cv f}$ may be the 
more appropriate measure when quantifying such asymmetric schemes.

We now discuss the physical significance of violating the field 
covariance inequality Eq.~(\ref{crit2}). It has been argued by some 
that our criteria are too stringent because the presence of 
entanglement in the teleporter can be demonstrated without exceeding 
our inequalities. For example the field covariance of a lossless, 
symmetric teleportation scheme which has an entanglement resource $V_{ent}$ 
($V_{ent}=1$ represents no entanglement whilst $V_{ent}\to 0$ 
represents maximal entanglement) and a gain of $\lambda$ (assumed 
real) is given by \cite{ral98,ral99}
\begin{equation}
V_{cv f}={{1}\over{2}}(1+\lambda)^{2}V_{ent}+{{1}\over{2}}
(1-\lambda)^{2}{{1}\over{V_{ent}}}
\end{equation}
Consider the case of unity gain ($\lambda=1$). Without entanglement 
($V_{ent}=1$) we find $V_{cv f}=2$. Only by introducing entanglement 
($V_{ent}<1$) can we obtain $V_{cv f}<2$. However it is not till we 
have introduced more than 50\% entanglement ($V_{ent}<.5$)
that we can obtain $V_{cv 
f}<1$ as per Eq.~(\ref{crit2}). Indeed it is in this intermediate region, 
where $2>V_{cv f}>1$, that the only experimental 
demonstration of continuous variable 
teleportation presently lies \cite{fur98}. One may ask if there is 
any {\it qualitative} difference between the types of correlation that 
can be observed when $2>V_{cv f}>1$ and those that can be observed 
when $1>V_{cv f}>0$. If qualitative differences exists then the 
more stringent definition of teleportation may be justified. We now 
give brief examples which illustrate that such differences do exist.

{\it Entanglement Requirements}. Firstly, it should be pointed out 
that true EPR entanglement is not required to reach the intermediate 
region. A single-mode squeezed beam, split on a 50:50 beamsplitter, is 
a sufficient resource \cite{loo992}. Although entangled \cite{aha66}, 
the beams so produced will exhibit 
non-classical correlations on only one quadrature with classical 
correlations on the conjugate quadrature. The signature of true EPR 
entanglement is non-classical correlations on both quadratures. The 
field covariance for a beam teleported with such a resource is given 
by \cite{ral00} 
\begin{equation}
V_{cv f}={{1}\over{4}}(1+\lambda)^{2}(1+V_{s})+{{1}\over{4}}
(1-\lambda)^{2}(1+{{1}\over{V_{s}}})
\label{sm}
\end{equation}
where $V_{s}$ is the level of single-mode squeezing in the resource. 
The value of Eq.(\ref{sm}) can never fall below 1 for any level of 
the squeezing or gain. Only when non-classical correlations can be 
observed on both quadratures can the field covariance fall below 1. 
Thus only with true EPR entanglement can we observe $V_{cv f}<1$.

{\it Preservation of Non-classical Statistics}. It is clearly of 
interest to ask under what conditions can teleportation of states 
with a non-classical character result in output states which retain 
that non-classical character. Suppose our input state is squeezed 
along the amplitude quadrature. We may ask under what conditions will 
the output still be squeezed. For symmetric, lossless teleportation 
the spectral variance of the output amplitude quadrature is given by 
\cite{ral98}
\begin{eqnarray}
V_{out}^{+} & = & {{1}\over{2}}(1+\lambda)^{2}V_{ent}+{{1}\over{2}}
(1-\lambda)^{2}{{1}\over{V_{ent}}}+\lambda^{2} V_{in}^{+}\nonumber\\
 & = & V_{cv f}+\lambda^{2} V_{in}^{+}
\label{ss}
\end{eqnarray}
It is immediately clear from Eq.(\ref{ss}) that it is impossible for 
squeezing to appear on the output unless $V_{cv f}<1$ is satisfied. In 
the limit of very strong squeezing ($V_{sq}^{+} \to 0$) squeezing on 
the output is guaranteed when the field covariance is less than 1, 
however in general a field covariance less than 1 is a neccessary but not 
sufficient condition for squeezing to be preserved.

{\it Preservation of Entanglement}. Of even greater importance for 
quantum information applications is to ask under what conditions 
entanglement between two systems, of which one has been teleported, 
is preserved. We have recently quantified this question by looking at 
the violation of a Bell-type inequality \cite{pol99}. In particular 
we looked at the value of the Clauser-Horne variable \cite{wal84,cla74}, 
$S$, between entangled photon beams when one of the beams is 
teleported using a continuous variable method. Local realistic hidden 
variable theories place the following restriction on the value of 
$S$; $S \le 1$. Quantum mechanical states allow $S$ to violate this 
inequality. The maximum violation occurs for non-maximally entangled 
states \cite{eb} and has the value; $S=1.5$ \cite{pol}.
Our result (in the limit of no loss) can be written
\begin{equation}
S={{{{1}\over{2}}(V_{cv f}-1)+\lambda^{2}(S_{i}+{{1}\over{2}})}\over{
(V_{cv f}-1)+2 \lambda^{2}}}
\label{se}
\end{equation}
where $S_{i}$ is the value of $S$ which would be obtained between the 
beams before teleportation. Eq.(\ref{se}) shows that if $V_{cv f}<1$ 
and $S_{i}=1.5$ then $S>1$. That is, provided that the entangled beams 
show a maximum violation of $S$ before teleportation, then some 
violation of local realism is guaranteed after teleportation if the 
field covariance falls below 1. As for squeezing this condition is 
neccessary, but not sufficient if $S_{i}$ does not have its maximum 
value.

We have derived the criteria for continuous variable teleportation 
from more fundamental arguments. We have shown by example that the 
properties that can be exhibited by the teleported system when the 
strong inequality $V_{cv f}<1$ is satisfied are qualitatively 
more ``quantum mechanical'' than in the region 
$2<V_{cv f}<1$. We have also noted that the strong inequality can only 
be satisfied by using true EPR entanglement in the teleporter.

This work was supported by the Australian Research Council.


\begin{thebibliography}{99}

\bibitem{ral98} T.~C.~Ralph and P.~K.~Lam, \prl {\bf 81}, 5668 (1998).

\bibitem{poi94} J.~-Ph.~Poizat, J.~-F.~Roch and P.~Grangier,
Ann.Phys.Fr. {\bf 19}, 265 (1993).

\bibitem{note1} Cross coupling between the quadratures can always be
eliminated by a suitable quadrature phase rotation.

\bibitem{loo99} P.~van Loock and S.~L.~Braunstein, LANL quant-ph/9902030 
(1999), S.~L.~Braunstein, C.~A.~Fuchs and H.~J.~Kimble, LANL
quant-ph/9910030 (1999).

\bibitem{art88} E.~Arthurs and M.~S.~Goodman, \prl {\bf 24}, 2447 (1988).

\bibitem{deu99} D.~Deutsch and P.~Hayden, LANL quant-ph/9906007 (1999).

\bibitem{hol90} M.~J.~Holland, M.~J.~Collett, D.~F.~Walls and
M.~D.~Levenson, \pra {\bf 42}, 2995 (1990).

\bibitem{cav82} C.~M.~Caves, \prd {\bf26},
1817 (1982).

\bibitem{ral99} T.~C.~Ralph, P.~K.~Lam and R.~E.~S.~Polkinghorne, 
J.Opt.B, {\bf 1}, 483 (1999).

\bibitem{fur98} A.~Furusawa, J.~L.~Sorensen, S.~L.~Braunstein, C.~A.~Fuchs, 
H.~J.~Kimble and E.~S.~Polzik, Science, {\bf 282}, 706 (1998).

\bibitem{loo992} P.~van Loock and S.~L.~Braunstein, \pra {\bf 61},
010302(R) (2000).

\bibitem{aha66} Y.~Aharonov et. al. Ann.Phys. (N.Y.) {\bf 39}, 498 
(1966).

\bibitem{ral00} T.~C.~Ralph, \pra {\bf 61}, 044301 (2000).

\bibitem{pol99} R.~E.~S.~Polkinghorne and T.~C.~Ralph, \prl {\bf 83}, 
2095 (1999).

\bibitem{wal84} D.~F.~Walls and G.~J.~Milburn, {\it Quantum Optics} 
(Springer-Verlag, Berlin, 1994).

\bibitem{cla74} J.~F.~Clauser and M.~A.~Horne, \prd
{\bf{10}}, 526 (1974).

\bibitem{eb} P.~H.~Eberhard, \pra {\bf 47} R747 (1993).

\bibitem{pol} R.~E.~S.~Polkinghorne, Department of Physics, 
University of Queensland, private communication (1999).

\end{thebibliography}
\end{document}